\documentclass{ws-procs9x6}

\begin{document}

\title{Critical $D$-module reps for finite superconformal algebras and their superconformal mechanics}

\author{F. TOPPAN}

\address{Teo Dep., CBPF, Rua Dr. Xavier Sigaud 150 (Urca),\\
Rio de Janeiro (RJ), cep 22290-180, Brazil\\
E-mail: toppan@cbpf.br}



\begin{abstract}
The simple finite Lie superalgebras $D(2,1;\alpha)$, $G(3)$, $D(4,1)$, $D(2,2)$, $A(3,1)$ and $F(4)$ admit $D$-module representations, given by a set of differential operators of a single variable $t\in \mathbb{R}$, at a critical value of the scaling dimension $\lambda$.
These superalgebras are one-dimensional ${\cal N}$-extended superconformal algebras with ${\cal N}=4$ ($D(2,1;\alpha)$), ${\cal N}=7$ ($G(3)$) and ${\cal N}=8$ (the remaining ones). The critical $D$-module reps induce invariant actions in the Lagrangian framework for superconformal mechanics in $D$ target dimensions. The ${\cal N}=8$ critical scalings $\lambda=\frac{1}{D-4}$ are linked to the $D$-dimensional harmonic functions with $D=1,2,3,5,6,7,8$. This talk is based on {J. Math. Phys.} {\bf 53} (2012) 043513 (arXiv:1112.0995), {J. Math. Phys.} {\bf 53} (2012) 103518 (arXiv:1208.3612)
and some extra material.
\end{abstract}



\section{Introduction}

The finite one-dimensional superconformal algebras are the simple Lie superalgebras \cite{{kac},{nrs},{dictionary}} ${\cal G}$
 which admit a grading
${\cal G} = {\cal G}_{-1}\oplus {\cal G}_{-\frac{1}{2}}\oplus
{\cal G}_0\oplus {\cal G}_{\frac{1}{2}}\oplus {\cal G}_{1}$
and whose even sector is ${\cal G}_{even}=sl(2)\oplus R$ (the subalgebra $R$ is known as $R$-symmetry). The odd sector (${\cal G}_{\frac{1}{2}}\oplus {\cal G}_{-\frac{1}{2}}$) is spanned by $2{\cal N}$ generators.\\
The positive sector ${\cal G}_{>0}$ is isomorphic to the algebra of the supersymmetric quantum mechanics \cite{witten} (at fixed ${\cal N}$).
Let $D,H,K$ be the $sl(2)$ generators; ${\cal G}_1$ (${\cal G}_{-1}$) is spanned by $H$ ($K$), while ${\cal G}_0=D{\Bbb C}\oplus R$.
Up to ${\cal N}\leq 8$ we have the following list of finite superconformal algebras over ${\mathbb C}$ (for ${\cal N}=0$ we recover the $sl(2)$ algebra):
\begin{eqnarray}\label{sca}
{\cal N}=0&:&  A_1,\nonumber\\
{\cal N} =1&:& B(0,1), \nonumber\\
{\cal N} =2&:& A(1,0), \quad C(2),\nonumber\\
{\cal N}=3 &:& B(1,1),\nonumber\\
{\cal N}=4 &:& A(1,1), \quad {\underline{D(2,1;\alpha)}},\nonumber\\
{\cal N}=5 &:& B(2,1),\nonumber\\
{\cal N}=6 &:&  A(2,1),\quad B(1,2), \quad D(3,1),\nonumber\\
{\cal N}=7 &:& B(3,1),\quad {\underline{G(3)}},\nonumber\\
{\cal N}=8 &:& D(4,1), \quad D(2,2), \quad A(3,1),\quad {\underline{F(4)}},
\end{eqnarray}
(the exceptional simple Lie superalgebras have been underlined). Their associated real forms can be found in [\cite{VP}].\par
A $D$-module rep of $sl(2)$, non-critical since it closes for any $\mu\in\mathbb{C}$, is given by
\begin{eqnarray}\label{sl2}
&H=\frac{d}{dt},\quad
D=-t\frac{d}{dt}-\mu, \quad
K=-t^2\frac{d}{dt} -2\mu t,
\end{eqnarray}
(with non-vanishing commutators $\relax [D,H]=H, [D,K]=-K, [H,K]=2D$).\par
In [\cite{{kt},{kht}}] the extensions of the  (\ref{sl2}) $D$-module rep to $D$-module reps (realized by a set of differential operators of a single variable $t\in \mathbb{R}$ acting on $n$ even and $n$ odd component fields) for the superconformal algebras with ${\cal N}>0$ were constructed.\par
They are based on the following two requirements: \\
{\em i}) a $D$-module rep of the ${\cal G}_>0$ subalgebra must coincide with a corresponding $D$-module rep of the global, one dimensional, ${\cal N}$-extended supersymmetry (based on [\cite{pt}], the latters have been classified, for ${\cal N}\leq 8$ and minimal number of bosonic and fermionic component fields, in [\cite{krt}]),\\
{\em ii}) the $sl(2)$ generators act diagonally (see [\cite{kt}]) on the component fields and $\mu$ is identified with their respective scaling dimensions.\par
For global ${\cal N}=4,8$ supersymmetries the minimal $D$-module reps are uniquely characterized by the assignment of the scaling dimensions of the component fields. They are specified by $(D,{\cal N}, {\cal N}-D)$, ($D$ fields of scaling dimension
$\lambda$, ${\cal N}$ fields of scaling dimension $\lambda+\frac{1}{2}$ and ${\cal N}-D$ fields of scaling dimension $\lambda+1$) (here $D=0,1,\ldots, {\cal N}$).
For ${\cal N}=7$ there exists a unique global supermultiplet which cannot be extended to a global ${\cal N}=8$ supermultiplet. Its ``field content" $(1,7,7,1)$ means that the assignment of the scaling dimensions is as follows: $1$ field of s.d. $\lambda$, $7$ fields of s.d.
$\lambda+\frac{1}{2}$, $7$ fields of s.d. $\lambda +1$, $1$ field of s.d. $\lambda+\frac{3}{2}$. \par
For ${\cal N}=4,7,8$ the (so far) arbitrary $\lambda$ is the overall scaling dimension of the global supermultiplet.\par
The construction of a $D$-module rep for the one-dimensional superconformal algebras is based on two inputs: a given $D$-module rep for the global ${\cal N}$-extended supersymmetry
algebra and the overall scaling dimension $\lambda$ (entering the $sl(2)$ subalgebra generators) of the supermultiplet. \par
Once identified these differential operators, there is no guarantee that their (repeated) commutators close producing a $D$-module rep of a finite superconformal algebra (to be identified). This is the case if a ``closure condition" is satisfied. The closure condition means, in particular, that the global supersymmetry generators of the ${\cal G}_{\frac{1}{2}}$ sector belong to a representation of the $R$-symmetry. In our investigation
three possibilities were encountered:
 \par
i) the closure condition is satisfied for any $\lambda$. A $D$-module rep of a finite superconformal algebra is obtained for any $\lambda$,\par
ii) the closure condition is never satisfied,\par
iii) the closure condition is satisfied for a critical value $\lambda=\lambda_{cr}$.

\section{The critical ${\cal N}=4,7,8$ $D$-modules.}

The minimal, global, ${\cal N}=4,7,8$ supermultiplets produce critical $D$-module representations of the corresponding superconformal algebras with the following identifications:
\par
${\cal N}=4$:  $D$-module reps are obtained for the $D(2,1;\alpha)$ superalgebras from the global $(D,4,4-D)$ supermultiplets, for any $\lambda$, with the identification $\alpha= (2-D)\lambda$,\par
${\cal N}=7$: a $G(3)$ $D$-module rep is obtained from $(1,7,7,1)$ at the critical value $\lambda= -\frac{1}{4}$,\par
${\cal N}=8$: $D$-module reps are obtained from $(D,8,8-D)$, $D\neq 4$, at the critical scalings
$\lambda_D = \frac{1}{D-4}$, with the identifications:
$D(4,1)$ for $D=0,8$,  $F(4)$ for $D=1,7$, $A(3,1)$ for $D=2,6$ and $D(2,2)$ for $D=3,5$.\par
Several comments are in order.  We list a few of them.\par
For other values of ${\cal N}$  (in particular ${\cal N}=2,3$) $D$-module reps with no criticality (closing for any value $\lambda$, were encountered).\par
The ${\cal N}=8$ global supermultiplet $(4,8,4)$ does not produce a $D$-module rep of an ${\cal N}=8$ superconformal algebra. \par
The ${\cal N}=7,8$ critical scaling dimensions are (partially) explained from the decomposition of the supermultiplets into two ${\cal N}=4$ supermultiplets, taking into account the relation between $\alpha$ and $\lambda$  and the isomorphism $D(2,1;\alpha)\equiv D(2,1;\alpha')$ for $\alpha, \alpha'$ related by an $S_3$-group transformation.\par
The $D$-module reps of the finite superconformal algebras allow to construct invariant actions (in Lagrangian framework) for superconformal mechanics (e.g., a uniquely defined $F(4)$-invariant action for the $(1,8,7)$ supermultiplet is produced \cite{{krt},{di},{kht}}, etc.).\par
For ${\cal N}=4$ the relation between $\alpha$ and $\lambda$ gives constraints on possible superconformal actions for interacting supermultiplets. In particular an admissible solution is found\cite{kht} for
$\alpha=\varphi$, the golden ratio, for the $D(2,1;\alpha)$-invariance of the $(1,4,3)$ and $(3,4,1)$ supermultiplets.

\section{Link between the ${\cal N}=8$ critical scalings and the harmonic functions}

A global ${\cal N}=8$-invariant action depends on a $D$-dimensional harmonic function\cite{{abc},{grt}} $\Phi$ ($\Box_D\Phi=0$). Its bosonic sector is a one-dimensional sigma model
${\cal S}= \int dt g_{ij}{\dot x}^i{\dot x}^j$ whose $D$-dimensional target metric is conformally flat ($g_{ij}=\delta_{ij} \Phi$), with $\Phi$ the conformal factor.\par
A conformal factor $\Phi=r^b$, dependent on the radial coordinate $r=\sqrt{\sum_i (x^i)^2}$
alone, is harmonic if the equation $b+D=2$ is satisfied. On the other hand,
the action ${\cal S}$ is scale invariant and contains no dimensional parameter if the scaling dimension $\lambda=[x^i]$ satisfies the condition $(b+2)\lambda -1+2=0$, where we have assumed, consistently with the conventions used in the previous Section, that $[t]=-1$. It is worth observing that, combining these two separate requirements, we recover the critical scaling dimensions $\lambda=\frac{1}{D-4}$ obtained from the closure condition of the $D$-module reps of the ${\cal N}=8$ superconformal algebras. Indeed, at the given target dimension $D$ and associated critical scaling we recover a superconformal mechanics invariant under the associated superconformal algebra. The scalar curvature $R$ of the target, computed from $\Phi$, is
$R=\frac{1}{4}r^{D-4}[(D-1)(D-2)^2(D-6)]$.\par
These results can be summarized in the table
\begin{eqnarray}
&
\begin{tabular}{|c|c|c|c|c|c|}\hline
$D$&$\Phi$&$R$&$\lambda_{cr}$&${\cal S}$&${\cal G}$\\ \hline
$0$&$-$&$-$&$-\frac{1}{4}$&$-$&$D(4,1)$\\ \hline
$1$&$r$&$0$&$-\frac{1}{3}$&$+$&$F(4)$\\ \hline
$2$&$1$&$0$&$-\frac{1}{2}$&$+$&$A(3,1)$\\ \hline
$3$&$r^{-1}$&$-\frac{3}{2}r^{-1}$&$-1$&$+$&$D(2,2)$\\ \hline
$4$&$r^{-2}$&$-6$&$-$&$-$&$-$\\ \hline
$5$&$r^{-3}$&$-18r$&$1$&$+$&$D(2,2)$\\ \hline
$6$&$r^{-4}$&$0$&$\frac{1}{2}$&$+$&$A(3,1)$\\ \hline
$7$&$r^{-5}$&$\frac{75}{2}r^3$&$\frac{1}{3}$&$+$&$F(4)$\\ \hline
$8$&$r^{-6}$&$126 r^4$&$\frac{1}{4}$&$+$&$D(4,1)$\\ \hline
\end{tabular}
&
\end{eqnarray}
Therefore superconformal mechanics can be constructed from $D$-module representations of superconformal algebras. See [\cite{fil}] and references therein for a recent review on superconformal mechanics and its applications (see also [\cite{ckp}] for a recent paper).

\section{Dual supermultiplets}

The ${\cal N}=8$ supermultiplets, both globally supersymmetric and superconformal, are dually related under the exchange $D\leftrightarrow 8-D$. In the superconformal case the duality gets manifested with the recovering of the same superconformal algebra (the critical scalings are related via the transformation $\lambda_D\leftrightarrow -\lambda_{8-D}$).\par
Let $\Psi\equiv\Psi_D$ denotes a superconformal multiplet with $8$ bosonic and $8$ fermionic component fields, while ${\widetilde \Psi}\equiv \Psi_{8-D}$ denotes a dual counterpart. A pairing $<\cdot|\cdot>$ exists such that the integral $\int dt <{\widetilde \Psi}|\Psi>$ is superconformally invariant. For a given superalgebra a generalized superconformal action
${\cal S}_{\alpha,\beta,\gamma}(\Psi,{\widetilde\Psi})$ can be given by setting ${\cal S}_{\alpha,\beta,\gamma}(\Psi,{\widetilde\Psi}) =\alpha{\cal S}_D(\Psi)+\beta{\cal S}_{8-D}({\widetilde \Psi}) +\gamma \int dt <{\widetilde \Psi}|\Psi>$, with
$\alpha,\beta,\gamma$ arbitrary non-dimensional constants and the ${\cal S}_D$, ${\cal S}_{8-D}$ actions given in Section {\bf 3}.\par
This classical superconformal action induces, in the path integral formulation, a superconformally invariant theory with partition function ${\cal Z}(\alpha,\beta,\gamma; J, {\widetilde J})$ ($J, {\widetilde J}$  are superconformal currents), given by
\begin{eqnarray}
{\cal Z}(\alpha,\beta,\gamma; J, {\widetilde J})&=&\int\int{\cal D}\Psi{\cal D}{\widetilde \Psi}
e^{-{{\cal S}_{\alpha,\beta,\gamma}(\Psi,{\widetilde\Psi})+\int dt <{\widetilde J}|\Psi>+\int dt <{\widetilde \Psi}|J>}}.
\end{eqnarray}
At least formally the measure ${\cal D}\Psi{\cal D}{\widetilde \Psi}$ is superconformally invariant.
Both the action and the measure are dimensionless. Indeed, the dimension $[{\cal D}\Psi]=
D\lambda_D-8(\lambda_D+\frac{1}{2})+(8-D)(\lambda_D+1) = 4-D$ is compensated by
$[{\cal D}{\widetilde \Psi}]=D-4$.
This construction naturally leads to superconformal models with double target manifolds.

\section{Conclusions and open problems}

The critical $D$-module representations induce invariant actions in the Lagrangian framework for classical superconformal mechanics. The quantization based on the functional approach via path integral is also naturally encoded. The superconformal invariance of the measure requires considering dually related supermultiplets leading to sigma-model actions with double target manifolds.\par
Several open questions, inspired by an operatorial quantization approach, can be raised. We are listing here a few of them.\par
A $D$-module rep induces a specific lowest weight representation (there exists a l. w. vector
$|w>$ s.t. $H|w>=0$ and $Q_i|w>=0$ for $i=1,\ldots, {\cal N}$) at the critical scaling dimension $\lambda$. Are the l.w.r.'s induced by a $D$-module rep unitary for some given real form of the $d=1$ SCA? Furthermore, is there an intrinsic, representation-theoretical characterization, to single out a l.w.r.
induced by a $D$-module, from a generic l.w.r. of the corresponding superalgebra?\par
It seems possible that unitary l.w.r.'s which are {\em not} induced by a $D$-module rep could define a quantum mechanics which does not admit a classical counterpart and a Lagrangian description.  \par
~\par
\verb|Acknowledgments|

\par
This work was supported by CNPq. The material here reported is partly based on a collaboration with Z. Kuznetsova and S. Khodaee.

\bibliographystyle{ws-procs9x6}
\bibliography{ws-pro-sample}

\end{document}